\documentclass{modelica}
\usepackage[utf8]{inputenc}
\usepackage{orcidlink}
\usepackage{bm}

\hypersetup{%
  pdftitle  = {Towards an FMI Layered Standard for DAE - Applications for Simulation and Optimization},
  pdfauthor = {Elmir Nahodovic, Andreas Heuermann, Joel A. E. Andersson, Adwait Verulkar, Srikanth Sivaramakrishnan, Masoud Najafi, Linus Langenkamp, Christian Bertsch, Torsten Sommer, Erik Henningsson, Hans Olsson, Bernhard Bachmann},
  pdfsubject = {American Modelica \& FMI Conference 2026},
  pdfkeywords = {FMI, Modelica, DAE, Differential Algebraic Equations, Layered Standard, Model Exchange, Dynamic Optimization},
  colorlinks,
  linkcolor=black,
  urlcolor=black,
  citecolor=black,
  pdfpagelayout = SinglePage,
  pdfcreator = latexmk}

\definecolor{xmltagcolor}{rgb}{0.13, 0.42, 0.65}
\definecolor{xmlattrcolor}{rgb}{0.50, 0.00, 0.50}
\definecolor{xmlcommentcolor}{rgb}{0.40, 0.40, 0.40}
\lstdefinelanguage{XML}{
  basicstyle=\footnotesize\ttfamily,
  morestring=[b]",
  morecomment=[s]{<!--}{-->},
  stringstyle=\color{stringcolor},
  commentstyle=\color{xmlcommentcolor}\itshape,
  identifierstyle=\color{xmltagcolor},
  morekeywords={valueReference,dependencies,index,version,encoding,xmlns,name,causality,variability,start,description},
  keywordstyle=\color{xmlattrcolor},
  showstringspaces=false,
}

\begin{document}
\thispagestyle{empty}

\title{Towards an FMI Layered Standard for DAE\\{\Large Applications for Simulation and Optimization}}
\author[1]{Elmir Nahodovic\,           \orcidlink{0009-0007-8738-3049}}
\author[2]{Andreas Heuermann\,         \orcidlink{0009-0000-1792-1701}}
\author[3]{Joel A.\ E.\ Andersson\,    \orcidlink{0009-0007-9206-3760}}
\author[4]{Adwait Verulkar\,           \orcidlink{0000-0002-2503-8457}}
\author[4]{Srikanth Sivaramakrishnan\, \orcidlink{0000-0001-9082-4182}}
\author[5]{Masoud Najafi\,             \orcidlink{0009-0006-2202-0266}}
\author[6]{Linus Langenkamp\,          \orcidlink{0009-0009-7517-4842}}
\author[7]{Christian Bertsch\,         \orcidlink{0009-0009-3627-0153}}
\author[1]{Torsten Sommer\,            \orcidlink{0009-0005-6424-4498}}
\author[1]{Erik Henningsson\,          \orcidlink{0000-0001-9881-0041}}
\author[1]{Hans~Olsson\,               \orcidlink{0009-0004-5370-9459}}
\author[6]{Bernhard Bachmann\,         \orcidlink{0000-0002-4339-0438}}

\affil[1]{Dassault Systèmes, Sweden, {\small\texttt{\{elmir.nahodovic, torsten.sommer, erik.henningsson, hans.olsson\}@3ds.com}}}
\affil[2]{Santa Anna IT Research Institute, Sweden, {\small\texttt{andreas.heuermann@santa-anna.se}}}
\affil[3]{FMIOPT AS, Norway, {\small\texttt{joel@fmiopt.com}}}
\affil[4]{General Motors, USA, {\small\texttt{\{adwait.verulkar, srikanth.sivaramakrishnan\}@gm.com}}}
\affil[5]{Siemens Industry Software, France, {\small\texttt{masoud.najafi@siemens.com}}}
\affil[6]{Bielefeld University of Applied Sciences and Arts, Germany, {\small\texttt{\{linus.langenkamp, bernhard.bachmann\}@hsbi.de}}}
\affil[7]{Bosch Research, Germany, {\small\texttt{christian.bertsch@de.bosch.com}}}

\maketitle\thispagestyle{empty}
\abstract{
The Functional Mock-up Interface (FMI) 3.0 standard for Model Exchange is restricted to hybrid ordinary differential equations, requiring any internal algebraic equations to be solved inside the Functional Mock-up Unit (FMU) before derivatives are returned to the importer.
For models originating from, e.g.\ Modelica, this means that non-linear algebraic equations must be solved through internal Newton iterations, which can reduce accuracy, increase computational cost, introduce hidden solver states, and cause robustness issues in downstream simulation and optimization workflows.
In this article, we present a proposal for a layered standard, \emph{fmi-ls-dae}, that exposes algebraic equations and their associated algebraic variables as part of a semi-explicit index-1 differential-algebraic equation.
We describe the proposed extensions to the FMI XML schema and demonstrate the approach through prototype implementations: Dymola and CasADi generate FMUs that expose this semi-explicit index-1 formulation, while CasADi, FMIOPT, Simcenter Twin Activate, fmusim, and MOO (the dynamic optimization tool of OpenModelica) import them for simulation and dynamic optimization.
On an industrially relevant multilink suspension corner model, the proposed DAE-FMU formulation enables the optimization routine to converge on an optimal control problem on which the equivalent ODE-FMU fails to converge.
We outline ongoing work towards consistent initialization, event handling and future support for higher-index DAEs.
}


\section{Introduction}
The Functional Mock-up Interface (FMI) version 3.0 for Model Exchange \cite{FMIv3:0:2} provides a standardized interface for models of dynamical systems in the form of Ordinary Differential Equations (ODEs) with state-machine event dynamics, often referred to as hybrid ODEs.
FMI 3.0 for Model Exchange allows an importer to simulate the model by supplying time, states, and inputs, and retrieving state derivatives and outputs.
However, this ODE-based interface imposes a restriction on the class of models that can be represented: any internal algebraic equations must be solved entirely inside the FMU before state derivatives are returned to the importer.

In Modelica, models are inherently described by Differential Algebraic Equations (DAEs).
During code generation, a Modelica tool performs symbolic transformations with the aim of reducing the high-index DAE \cite{ModelicaSpec}.
Usually it is reduced to an index-1 DAE, which is then typically converted into an ODE by solving the algebraic equations at each simulation step \cite{cellier2006continuous}.
Some tools, such as OpenModelica and Dymola, also support a \emph{DAE-Mode}, where the semi-explicit index-1 DAE is instead passed directly to an appropriate implicit solver such as IDA or Radau IIA. Rather than being solved separately, the non-linear equations are then solved along with the non-linear equations of the implicit integrator method, which can improve simulation time of Modelica models containing large systems of non-linear equations \cite{BraunOpenModelicaDAE, Henningsson2019DAE}.
However, such a DAE simulation is currently not possible if the model is exported as a Functional Mockup Unit (FMU) adhering to FMI~3.0 for Model Exchange, since the standard only supports ODEs.

Recent work has also shown growing interest in using Model Exchange FMUs directly within optimization workflows, in particular with FMUs exported from Modelica models (e.g., \cite{Pfeiffer2025}).
Optimization problems involving FMUs, including trajectory optimization problems and steady-state optimization problems, are typically solved iteratively with gradient-based methods.
In such formulations, it is often preferable to handle the algebraic equations at the optimization level rather than inside the FMU.

In this article, we show how to expose algebraic equations during continous time simulation of an FMU in a standardized way.
This standardizes existing tool-internal DAE-Mode so it can be used across the FMI ecosystem, allowing an importing tool with support for DAEs to use them more efficiently.
This work uses the concept of \emph{layered standards}, which was developed by the Modelica Association to enable extensions for new use cases while maintaining compatibility with the core standard~\cite{Bertsch2023}. This article covers the \texttt{1.0.0-alpha.1}\footnote{
    GitHub: modelica/fmi-ls-dae milestone \href{https://github.com/modelica/fmi-ls-dae/milestone/1}{Alpha-1}
} release, planned for 2026.

We present prototypes demonstrating how exposing non-linear algebraic equations through a Layered Standard for FMI Model Exchange can substantially improve the precision and robustness of optimization and simulation tasks involving such FMUs.
A prototype with Dymola and CasADi generates FMUs that include this additional information.
On the import side, the usability of the FMI-based Layered Standard for DAEs is demonstrated with prototypes in CasADi, Simcenter Twin Activate, fmusim, and MOO, the dynamic optimization tool used by OpenModelica.

\section{A Layered Standard for DAEs}
The proposed layered standard for differential algebraic equations, \emph{fmi-ls-dae}, extends FMI 3.0 to support the exchange of dynamic models in DAE form \cite{FMI-LS-DAE-2026}.

The concept of layered standards on top of the FMI standard was introduced to keep the core FMI standard lean and small, while extending it for new use cases.
The FMI standard has certain extension mechanisms for layered standards, such as annotations, a defined subfolder in the FMU for additional files, and a manifest XML file.
But even with the additional optional DAE support, the FMU shall still be a valid Model Exchange FMU that can be simulated in all tools with FMI capability without using the additional features of the new layered standard.

This article focuses on the specific use case of exposing systems of non-linear algebraic equations during continuous-time integration, such that a single FMU represents an index-1 semi-explicit DAE.
This is achieved by augmenting the active equations during \emph{Continuous-Time Mode} to include the residual equations of the algebraic constraints.

We will briefly discuss remaining work for the first release and the considerably broader scope of fully supporting higher-index hybrid DAEs in \autoref{sec:outlook}.

\subsection{Previous work on DAE support in FMI}

Support for Differential Algebraic Equations in FMI has been discussed in the FMI Project for a long time.
In 2015--2018 a working group was formed that collected requirements, discussed different approaches, and created a first internal draft.
At that time, the concept of Layered Standards for FMI did not yet exist, so an optional extension of the FMI for Model Exchange was discussed for a version v2.1 or later v3.0 of the FMI standard.
The main motivation at that time was the modular exchange and coupling of simulation models that are better described as DAEs (e.g., multibody systems).
However, for the final FMI 3.0 feature set it was decided not to include DAE support, as there were unresolved questions, with respect to
initialization at the start of the simulation, and at events.
The current proposal is based on this previous work and extends it, but is motivated mainly by new use cases such as optimizing FMUs.

\subsection{Mathematical Formulation}
We consider a semi-explicit index-1 DAE in state-space form:
\begin{equation}
\begin{aligned}
    \dot{x} &= f(x,\, w,\, u,\,t) \\
    0       &= g(x,\, w,\, u,\,t) \\
    y       &= h(x,\, w,\, u,\,t),
\end{aligned}
\label{eq:ind1-dae}
\end{equation}
where $x$ are the continuous-time states, $w$ the algebraic variables, $u$ the inputs, $t$ the independent variable (time), and $y$ the outputs.
The functions $f$, $g$, and $h$ define the state derivatives, residuals, and outputs, respectively.
We assume that $\frac{\partial g}{\partial w}$ is non-singular, so that $w$ is defined implicitly by $g$.
In this article, $g$ represents the system of non-linear equations to be exposed, with $w$ as the iteration variables of those systems.
We emphasize that the index-1 property ensures that $\frac{\partial g}{\partial w}$ is structurally non-singular (i.e., the system is well-posed in principle),
but does not guarantee numerical non-singularity.
Near-singular Jacobians can still occur in practice due to physical operating points or parameter regimes.

\subsection{Standard Model Exchange with Hybrid ODEs}
\begin{figure}[htb]
    \centering    \includegraphics[width=0.8\columnwidth]{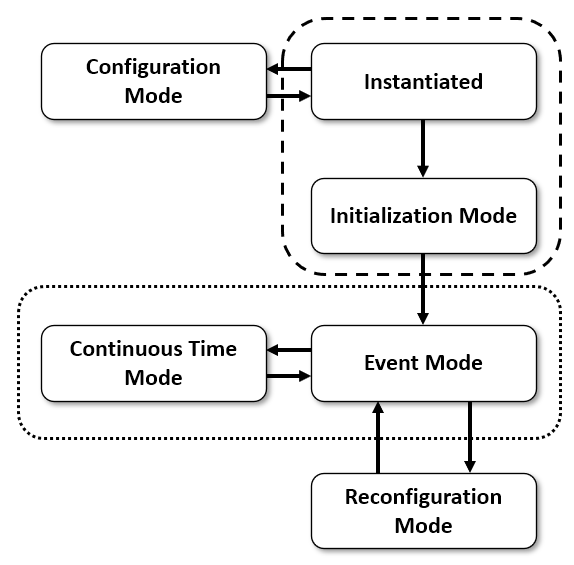}
    \caption{Simplified State Machine for Model Exchange with FMI 3.0.
        The dashed and dotted lines indicate \emph{Super State}s of the FMU.
        The dashed line around \emph{Instantiated} and \emph{Initialization Mode} is \emph{FMU State Settable} while the dotted line around \emph{Continuous-Time Mode} and \emph{Event Mode} is \emph{Initialized}.
        The modes \emph{Configuration}- and \emph{Reconfiguration Mode} allow the importer to set \texttt{structuralParameter}s and \texttt{tunable} \texttt{structuralParameter}s \cite[see Section~3.1, State Machine for Model Exchange]{FMIv3:0:2}.}
    \label{fig:state-machine}
\end{figure}

The Model Exchange interface is governed by a state machine in which different sets of equations become active depending on the current \emph{mode}, see \autoref{fig:state-machine} giving a simplified view.

Three of the modes presented in \autoref{fig:state-machine} contain governing equations that are of particular interest when expanding to DAEs.
We give a simple overview of these modes:

The \emph{Initialization Mode} establishes consistent initial values for all variables before the simulation begins.

The \emph{Event Mode} is active whenever a discrete change occurs, such as a state or time event.
Here the discrete-time equations are evaluated and the discrete states are updated.
The FMU may also reinitialize continuous states after an event.

And finally, the \emph{Continuous-Time Mode} is the integration phase between events.
A simplified view is that the importer sets the current time $t$, the continuous inputs $u$, and the continuous states $x$, using the returned state derivatives $\dot{x}$ and outputs $y$ evaluated by the FMU.

The current FMI 3.0 Model Exchange standard supports hybrid ODEs only.
For the DAE formulation of \autoref{eq:ind1-dae}, this means that the algebraic variables $w$ and the corresponding equations $g(x, w, u, t) = 0$ are treated as internal to the FMU.
By the implicit function theorem, provided $\partial g / \partial w$ is non-singular (as is the case for an index-1 system), $w$ can be locally eliminated and written as $w = g^{-1}(x, u,t)$, usually evaluated iteratively using some variant of Newton's method when $g$ consists of systems of non-linear equations.
Consequently, the equations governing Continuous-Time Mode that are exposed through the FMU C-API do \emph{not} represent $f(x, w, u, t)$ and $h(x,w,u,t)$ from \autoref{eq:ind1-dae}, but the composed functions $f(x, g^{-1}(x, u,t), u,t)$ and $h(x, g^{-1}(x, u,t), u,t)$, with the algebraic structure hidden from the importer.
In regular Model Exchange, the importer thus sees the following equations during Continuous-Time Mode:

\begin{equation}
\begin{aligned}
    \dot{x} &= f(x, g^{-1}(x, u, t), u,\,t) \\
    y       &= h(x, g^{-1}(x, u, t), u,\,t).
\end{aligned}
    \label{eq:me-cont-ginv}
\end{equation}
The variables representing $\dot{x}$ and $y$ in the FMU can be acquired by calling \texttt{fmi3Get\{VariableType\}} which will evaluate the functions $f, g^{-1},$ and $h$ in \eqref{eq:me-cont-ginv}.
The evaluation of $g^{-1}$ in \eqref{eq:me-cont-ginv} typically requires a Newton iteration inside the FMU.

Internal Newton iterations can be problematic for multiple reasons, as discussed below.
\subsubsection*{Lower accuracy}
The internal Newton iteration lowers accuracy.
While model equations without algebraic loops can often be evaluated to machine precision, an iterative solver is limited to the stopping criteria of the method, often lower order.
This inaccuracy can lead to convergence issues for example in an outer optimization problem, which is typically also solved iteratively using a Newton-type method with similar tolerances.

\subsubsection*{Higher computational cost}
Eliminating algebraic variables leads to nested non-linear solves and prevents exploiting the coupled Jacobian structure, often resulting in higher computational cost compared to a single Newton solve in a DAE formulation.

\subsubsection*{Hidden state inside the solver}
Solving algebraic variables internally introduces hidden solver state (e.g., initial guesses, iteration history), making function evaluations path-dependent and potentially non-reproducible.

Exposing algebraic variables in a DAE formulation avoids this issue and improves robustness, in both optimization and simulation contexts.

\subsubsection*{Failure to Converge}
In a direct DAE solver, if the algebraic constraint is hard to meet, the solver naturally reduces the time step ($\Delta t$) to help the global Newton iteration converge.
In the nested approach: The inner $w$-solver has no knowledge of the ODE's time step.
If the inner solver fails to find $w$ (because the initial guess from the last step was too far off), the whole simulation may fail.
It is much harder to implement robust ``error recovery'' when the failure happens inside a black-box function evaluation.

\subsubsection*{Singularity of $g^{-1}$ and Near-Singular Jacobians}
In the current \emph{fmi-ls-dae} standard, only semi-explicit index-1 DAEs are considered.
The index-1 assumption requires $\frac{\partial g}{\partial w}$ to be locally non-singular, allowing the algebraic variables to be
defined implicitly through the algebraic constraints.
In practice, however, $\frac{\partial g}{\partial w}$ may become poorly conditioned near physically relevant operating points,
such as bifurcations, parameter boundaries, or stiff regimes.

The standard itself does not prescribe a specific treatment for such situations.
Rather, by exposing the coupled DAE system

\[
\mathbf{J}_{\mathrm{DAE}} =
\begin{pmatrix}
\frac{\partial f}{\partial x} & \frac{\partial f}{\partial w} \vspace{.1cm} \\
\frac{\partial g}{\partial x} & \frac{\partial g}{\partial w}
\end{pmatrix},
\]

to the simulation environment, it enables dedicated DAE solvers to apply robustness measures such as adaptive step-size control, scaling, regularization,
 and consistent initialization.
 In contrast, the nested ODE formulation relies on the local elimination of the algebraic variables, for which the sensitivity

\[
\frac{\partial w}{\partial x}
=
-\left(\frac{\partial g}{\partial w}\right)^{-1}
\frac{\partial g}{\partial x}
\]

is directly affected by the conditioning of $\frac{\partial g}{\partial w}$.
Consequently, numerical errors may be significantly amplified when $\frac{\partial g}{\partial w}$ is nearly singular.

Although the DAE formulation does not eliminate singularity-related difficulties, it avoids the explicit construction of the reduced dynamics and gives the solver direct access to the coupled system, generally improving the detection and management of ill-conditioned situations \cite{Brenan1989,Hairer1991}.
Cases in which $\frac{\partial g}{\partial w}$ becomes singular correspond to a local breakdown of the index-1 assumption and are outside the scope of the current version of the standard, remaining a topic for future extensions.

\subsection{Extensions via \emph{fmi-ls-dae}}
The core idea of \emph{fmi-ls-dae} is to augment the active equations during Continuous-Time Mode (\autoref{eq:me-cont-ginv}) with \emph{residual} variables $r$ and \emph{algebraic variables} $w$:
\begin{equation} \label{eq:ls-dae-cont-g}
\begin{aligned}
\dot{x} &= f(x,\, w,\, u,\,t) \\
r       &= g(x,\, w,\, u,\,t) \\
y       &= h(x,\, w,\, u,\,t),
\end{aligned}
\end{equation}
to properly represent the mathematical formulation of \autoref{eq:ind1-dae}.
It is the responsibility of the importer to find values for $w$ such that the residuals $r$ are zero.

The layered standard adds an FMI-LS-DAE XML manifest file which tags the DAE related variables from the \verb|ModelVariables| element in the modelDescription.xml, and adds a new model structure which contains residual equations (the constraints), available in the \verb|extra/| directory \cite[see Section~2.6. Versioning and Layered Standards]{FMIv3:0:2}.

In the current version, there is no extension of the FMI C API.

\subsubsection{DAE-Mode}
\label{sec:dae-mode}
A layered standard requires that any importer not supporting this extension can simply ignore that semantic and simulate the Model Exchange FMU as a hybrid ODE.
This means that an importer should ``opt-in'' to simulate the FMU in DAE-Mode with the residuals exposed to be solved by the importer.
To handle this we propose that an FMU supporting \emph{fmi-ls-dae} must expose a structural parameter called \emph{enableDAEModeParameter} of type \texttt{Boolean} with variability \texttt{fixed} or \texttt{tunable} and start set to \texttt{false}, as shown in \autoref{code:DAEMode_xml}.

\begin{lstlisting}[float, language=XML, caption={Structural parameter \emph{enableDAEModeParameter} exposed by an FMU to opt in to DAE-Mode as defined by \emph{fmi-ls-dae}.}, label = {code:DAEMode_xml}]
<Boolean name="<free to choose>"
 valueReference="<free to choose>"
 causality="structuralParameter"
 variability="tunable|fixed"
 start="false"
 description="Set to true to enable DAE mode
              as defined by FMI-LS-DAE."/>
\end{lstlisting}

This parameter can be set in \emph{(Re)Configuration Mode} (see \autoref{fig:state-machine} for the simplified view of the state machines of the FMU); setting it to \texttt{true} activates the equations in \eqref{eq:ls-dae-cont-g}.

\subsubsection{Initialization}
\label{sec:initialization}
Initialization in DAE mode is not yet fully specified, since the first alpha release focues on exposing the DAE formulation during Continuous-Time Mode. Nevertheless, this section details the current ideas regarding initialization.

Computing a \emph{consistent} initial state $\left( x(t_0), w(t_0) \right)$ using

\begin{equation}
    \label{eq:alg-constraint}
    0 = g(x(t_0), w(t_0), u(t_0), t_0)
\end{equation}

\noindent
and initial conditions is a non-issue for the importer of the FMU in ODE mode, since the FMU will handle it internally.
When \emph{enableDAEModeParameter} is \texttt{false} during initialization the FMU will be initialized in ODE-Mode.
After the Initialization Mode is exited the FMU is in Event Mode and allowed to enter Reconfiguration Mode to switch into DAE-Mode.

Another option is to switch into DAE-Mode during Configuration Mode and delegate the computation of a consistent initial condition to the importing tool, which may apply suitable numerical or symbolic methods, or rely on expert-supplied initial guesses.
In the most direct case, the user provides a fully consistent set of initial values $\left( x(t_0), w(t_0) \right)$.
When start values are provided only for the differential states $x(t_0)$, initialization reduces to solving the algebraic constraint \eqref{eq:alg-constraint} for $w(t_0)$ alone via Newton's method.

\subsubsection{Algebraic variables}
The algebraic variables $w$ are specified by the \verb|<AlgebraicVariables>| element in the manifest file, referencing existing entries in the model description XML element \verb|<ModelVariables>| by their \texttt{valueReference}.

For the C-API this means that the FMU must allow the importer to call \texttt{fmi3Set\{VariableType\}}, which is normally not permitted for \texttt{local} variables.
Moreover, they must have the possibility to be exposed as an \verb|<InitialUnknown>| in the new \verb|<ModelStructure>| (see section \autoref{sec:ModelStructure}), since the initial equations defined in Initialization Mode now also may initialize the algebraic variables.

\subsubsection{Model Structure} \label{sec:ModelStructure}
A DAE-version of element \verb|<ModelStructure>|, extended with \verb|<Residual>| elements is added to the FMI-LS-DAE XML manifest file.
Each \verb|<Residual>| encodes the constraint $r$ from \eqref{eq:ls-dae-cont-g}, and carries a \verb|valueReference| identifying the residual variable, an optional \texttt{dependencies} list of known value references, and an optional \texttt{dependenciesKind} list.

For the purpose of exposing the semi-explicit index-1 formulation in \eqref{eq:ind1-dae} it should be assumed that the DAE formulation represented by the new \verb|<ModelStructure>| defines an index-1 system. Including higher-index formulations are planned and discussed in section \ref{sec:outlook-higher-index}.

The \verb|dependencies| attribute of the \verb|<Residual>|, \verb|<Output>|, and \verb|<ContinuousStateDerivative>| may include algebraic variables.

\subsubsection{Example}
To illustrate the extensions of the XML schema, we will be studying a simple academic DAE.
It is described by 6 scalar equations:

\begin{equation}
\label{eq:simpleDAE}
\begin{aligned}
    \dot{x_1} &= \sin(x_1) + \sin(w_1 w_2 u_1)               \\
    \dot{x_2} &= \sin(x_1 x_2) + \sin(w_1 w_2 u_2) + u_1^2   \\
    0         &= w_1 u_1^2 + \tanh(3 w_1) + u_2 x_1^3        \\
    0         &= \frac{\exp(w_1 w_2 u_1)}{3} - \sin(w_2 x_2) \\
    y_1       &= u_1 x_1 w_1 + \sin(u_2)                     \\
    y_2       &= u_2 x_2 w_2 + \sin(u_1 u_2)
\end{aligned}
\end{equation}

Here $x_1, x_2$ are state variables with state derivatives $\dot{x_1}, \dot{x_2}$, $w_1, w_2$ are algebraic variables, $u_1, u_2$ inputs and $y_1, y_2$ outputs.
Initial conditions are provided by $x_1(0) = 0.5$ and $x_2(0) = 0.5$.
In this case $f$, $g$, and $h$ are represented by the vectors
\begin{equation}
\begin{aligned}
    f &= \begin{bmatrix} \sin(x_1) + \sin(w_1 w_2 u_1) \\ \sin(x_1 x_2) + \sin(w_1 w_2 u_2) + u_1^2 \end{bmatrix}, \\
    g &= \begin{bmatrix} w_1 u_1^2 + \tanh(3 w_1) + u_2 x_1^3 \\ \frac{\exp(w_1 w_2 u_1)}{3} - \sin(w_2 x_2) \end{bmatrix}, \\
    h &= \begin{bmatrix} u_1 x_1 w_1 + \sin(u_2) \\ u_2 x_2 w_2 + \sin(u_1 u_2) \end{bmatrix}.
\end{aligned}
\end{equation}

Consider an FMU of \autoref{eq:simpleDAE}, with a \verb|valueReference| for each variable according to \autoref{tab:valueReferences}.
The proposed element listing the algebraic variables, as well as extended \verb|<ModelStructure>| for Continuous-Time Mode is presented in \autoref{code:AlgebraicVariables_xml} and \autoref{code:ModelStructure_xml} respectively:
\begin{table}[h]
    \centering
    \caption{Value references for the variables in \eqref{eq:simpleDAE}. The variables $r_1$ and $r_2$ represent the residuals for the non-linear equations defined by $g$.}
    \label{tab:valueReferences}
    \begin{tabular}{l c}
        \toprule
        Variable & valueReference \\
        \midrule
        $x_1$       & 0  \\
        $x_2$       & 1  \\
        $\dot{x_1}$ & 2  \\
        $\dot{x_2}$ & 3  \\
        $w_1$       & 4  \\
        $w_2$       & 5  \\
        $u_1$       & 6  \\
        $u_2$       & 7  \\
        $y_1$       & 8  \\
        $y_2$       & 9  \\
        $r_1$       & 10 \\
        $r_2$       & 11 \\
        \bottomrule
    \end{tabular}
\end{table}
\begin{lstlisting}[language=XML, caption={The two algebraic variables from \autoref{eq:simpleDAE} are exposed through a new element in the fmi-ls-manifest.xml.}, label = {code:AlgebraicVariables_xml}]
<AlgebraicVariables>
    <AlgebraicVariable valueReference="4"/>
    <AlgebraicVariable valueReference="5"/>
</AlgebraicVariables>
\end{lstlisting}
\begin{lstlisting}[language=XML, caption={Example for the extended ModelStructure (for Continuous-Time Mode) for \autoref{eq:simpleDAE}}, label = {code:ModelStructure_xml}]
<ModelStructure>
  <Output
    valueReference="8"
    dependencies="0 4 6 7"/>
  <Output
    valueReference="9"
    dependencies="1 5 6 7"/>
  <ContinuousStateDerivative
    valueReference="2"
    dependencies="0 4 5 6"/>
  <ContinuousStateDerivative
    valueReference="3"
    dependencies="0 1 4 5 6 7"/>
  <Residual
    valueReference="10"
    dependencies="0 4 6 7"/>
  <Residual
    valueReference="11"
    dependencies="1 4 5 6"/>
</ModelStructure>
\end{lstlisting}
All existing elements of the \verb|<ModelStructure>| in \autoref{code:ModelStructure_xml} above retain their standard semantics; the only change is that their \texttt{dependencies} attributes may now also reference algebraic variables, and that the algebraic variables can be listed among the \verb|<InitialUnknown>|.

It should be emphasised that the \verb|dependencies| and \verb|dependenciesKind| attributes will differ between ODE- and DAE Mode in more ways than just including \verb|valueReference|s of the algebraic variables.
For example with the SimpleDAE in \autoref{eq:simpleDAE}, if it were represented as an ODE, then the derivative $\dot{x}_1$ will implicitly depend on the input $u_2$ through the algebraic variables appearing in the residual equations.
However, in the DAE formulation these algebraic variables appear directly in $\dot{x}_1$, and the input dependencies are instead encoded in the residual equations.

\subsubsection{Partial derivatives}
The standard directional and adjoint derivative APIs (\verb|fmi3GetDirectionalDerivative|, \verb|fmi3GetAdjointDerivative|) are used as defined in the FMI 3.0 core standard, extended so that algebraic variables may appear as knowns for any unknown, and residual variables may appear as unknowns.

\section{Implementations}
This section details prototype implementations developed alongside the proposals described in this article.
It should be noted that \emph{fmi-ls-dae} has not yet released any formal specification for the semi-explicit index-1 formulation; what is presented here reflects the authors' current proposals along with the \emph{fmi-ls-dae} working group's discussions, and is subject to change as the standard evolves.

\subsection{Dymola}
Dymola (Dynamic Modeling Laboratory) is a commercial tool for modeling and simulation based on Modelica and FMI.
A Modelica model is in general described by a high-index implicit DAE:
\begin{equation}
\begin{aligned}
    0 &= F(\dot{x}, x, w, u, t, d) \\
    d &= \alpha(\dot{x}, x, w, u, t, \texttt{pre}(d)) \\
    y &= H(\dot{x}, x, w, u, t, d).
\end{aligned}
\label{eq:dymola-implicitdae}
\end{equation}
Where $x$ are the dynamic variables with derivatives $\dot{x}$, $w$ are algebraic variables, and $t$ is time.
Here $d$ are the discrete variables, which we ignore for now.
The variables $u$ and $y$ represent any top-level inputs and outputs.
During code generation, a Modelica tool applies symbolic transformations to reduce \autoref{eq:dymola-implicitdae} to a form suitable for simulation.
Algebraic variables that do not affect the dynamics (auxiliary variables) are eliminated in this process.
Loops that can be solved symbolically (such as linear systems or inversions of elementary functions like $\exp$ and $\sin$) are solved at code generation time.
What remains are the irreducible non-linear systems of equations that affect the differentiated variables, where the non-linear equations depend on the variables that appear differentiated ($x$) and iteration variables ($w$), which together with the dynamic equations yield the index-1 semi-explicit DAE of \eqref{eq:ind1-dae}.
It is precisely this form of semi-explicit DAE with $g$ consisting of non-linear equations that are exposed with the generated DAE-FMU.

\subsubsection{Generating the FMU}

As of the release of Dymola 2026x Refresh 1 on April 17, 2026, to support the layered standard for Differential Algebraic Equations, Dymola gives the option to expose these non-linear equations and iteration variables in the FMU, to be handled by the importer.
The non-linear equations (residuals) are exposed as variables in the FMU with the algebraic variables tagged in the \texttt{fmi-ls-manifest.xml}. The partial derivative functions \texttt{fmi3GetDirectionalDerivative} and \texttt{fmi3GetAdjointDerivative} have been updated to take in residuals as unknowns and algebraic variables as knowns.

Note that after exiting Initialization Mode, the FMU will initialize the iteration variables to solve the residuals, then during Continuous-Time Mode, it is the importer's responsibility to solve the residual equations.
Therefore, the DAE FMUs generated from Dymola 2026x Refresh 1 are automatically not valid ODE FMUs since the \emph{enableDAEModeParameter} hasn't been implemented at the time of that release.

One current limitation is that models with events exported as FMUs do not work in general, since there is not yet a formal proposal for extending the Model Exchange interface to handle discrete behavior in hybrid DAEs.
Consequently, dynamic state selection is also not supported for DAE-FMUs yet, as the selection of states is generally discontinuous.

At the time of writing, all the above limitations regarding \emph{enableDAEModeParameter}, events, and dynamic state selection are being implemented in a future Dymola release.

\subsection{CasADi}  \label{sec:casadi}
CasADi \cite{Andersson2019} is an open-source framework for C++, Python, MATLAB and Octave.
In particular, CasADi provides building blocks for efficiently formulating and solving non-linear optimization problems, specifically problems involving dynamic systems.
CasADi has been used to implement various optimization algorithms in industry and academia and is used in the backend of multiple open-source and commercial tools for dynamic optimization.

Since 2023, CasADi allows standard Model Exchange FMUs to be imported and used as twice differentiable functions within the framework \cite{Andersson2023}.
At the same time, FMI~3.0 export was added to CasADi, allowing users to export models formulated into CasADi as standard FMUs.
CasADi generated FMUs include serialized CasADi expressions via layered standard, providing a standard mechanism to recover the original CasADi expressions from the FMU when imported back into CasADi.

As part of this work, CasADi's FMU import and export functionality was extended to DAEs, via the \emph{fmi-ls-dae} layered standard.

\subsection{FMIOPT} \label{sec:fmiopt}
FMIOPT is a commercial, general-purpose dynamic optimization framework built on top of CasADi.
It aims at providing a robust and efficient workflow for solving dynamic optimization problems, starting with challenging simulation models.
Its method-agnostic and FMI-based middle-layer allows the numerical solution to be performed in ephemeral solver instances in high-performance parallelized hardware, including in a cloud environment.
Features include multiple sequential and simultaneous solution methods, minimal-time formulations, multi-stage formulations, visualizations and advanced diagnostics.

\subsection{Simcenter Twin Activate}

Simcenter Twin Activate is a general-purpose modeling and simulation platform designed for hybrid dynamical systems.
It supports ordinary differential equations (ODEs) and fully-implicit index-1 and semi-explicit index-2 differential-algebraic equations (DAEs), along with ODEs combined with constraints (invariants).
The software includes multiple numerical solvers for both ODE and DAE systems.

Historically, many models resulted in high-index DAEs that could not be exported as Model Exchange FMUs.
The only viable option was to export them with FMI for Co-Simulation, which performed poorly when connected to other FMUs.
The introduction of DAE-based dynamical systems in the Model Exchange FMU standard has opened new possibilities, allowing high-index models to be exported as FMUs compatible with other vendors' tools.

To validate this capability, Simcenter Twin Activate has enhanced its FMU import functionality to support the new DAE features and imported and tested two DAE-FMUs in \autoref{sec:simcenter_experiment}.

\subsection{MOO}

\label{sec:moo_intro}
MOO (Modelica / Model Optimizer) is a library for dynamic optimization and forms the basis for the new optimization runtime of OpenModelica.
The presented FMI interface prototype extends MOO to support dynamic optimization of FMUs on a fixed time horizon $[t_0, t_f]$.
The supported problem class is given by
\begin{equation}
	\begin{aligned}
		\min_{u, p}\quad  &M(x_0, w_0, x_f, w_f) + \int_{t_0}^{t_f} L\bigl(x, w, u, t\bigr)\, dt \\
		\text{s.t.}\quad  \dot{x} &= f\bigl(x, w, u, t\bigr) \\
		                  0 &= g(x, w, u, t) \\
		                  c^L &\le c\bigl(x, w, u, t\bigr) \le c^U \\
		                  \psi^L &\le \psi\bigl(x_0, w_0, x_f, w_f\bigr) \le \psi^U
	\end{aligned}
\end{equation}
The objective consists of a Mayer term $M$ and a Lagrange term $L$.
Dynamics are described by the differential equations $\dot{x} = f(\cdot)$ with states $x$, controls $u$, and explicit algebraic variables $w$ with associated algebraic constraints $g(\cdot)=0$.
Path and boundary constraints are defined by the functions $c$ and $\psi$, respectively.
MOO additionally supports static parameter optimization and free initial and final times.
These features could be integrated into the FMI workflow by exposing FMU parameters and time variables as optimization variables, but are not yet part of the current prototype.

In the current prototype, differential and algebraic equations are identified from the FMU, while selected FMU value references can be used to define the Mayer and Lagrange
terms, as well as path and boundary constraints with their corresponding bounds.
The interface provides the required function evaluations and Jacobians to the optimizer.
Second-order derivatives are currently not available, but could be provided using finite differences.
The prototype is available on GitHub\footnote{
    GitHub repository: \href{https://github.com/AMIT-HSBI/MOO/tree/fmi-ls-dae}{AMIT-HSBI/MOO} branch fmi-ls-dae
}.

\subsection{fmusim}
fmusim\footnote{
    GitHub repository: \href{https://github.com/modelica/fmusim}{https://github.com/modelica/fmusim}
} is a command line tool to import, validate, and simulate Functional Mock-up Units, written in Rust.
fmusim has been expanded with the option to import FMUs conforming to \emph{fmi-ls-dae} and supplies the Sundials IDA solver \cite{gardner2022sundials, hindmarsh2005sundials} for its simulation of such Model Exchange FMUs.

\section{Examples} \label{sec:examples}
Four DAE-FMUs were exported using Dymola 2026x Refresh 1:

\begin{itemize}
    \item \texttt{SimpleDAE.fmu}:
        Based on \autoref{eq:simpleDAE}.

    \item \texttt{Fourbar1.fmu}:
        Based on Modelica model \texttt{Modelica.\allowbreak Mechanics.\allowbreak MultiBody.\allowbreak Examples.\allowbreak Loops.\allowbreak Fourbar1} from the Modelica Standard Library version 4.1.0.
        A simple kinematic loop consisting of 6 revolute joints, 1 prismatic joint and 4 bars that is often used as basic constructing unit in mechanisms - was exported as \texttt{Fourbar1.fmu}.

    \item \texttt{Crane.fmu}: Describes a crane with mass $m_C$, actuated by a force $u$ and carrying a load with mass $m_L$ suspended by a rope of length $l$:
\begin{equation}
\begin{aligned}
  m_L \, \ddot{x_L} &= \frac{x_L-x_C}{l} \, T, \qquad
  m_L \, \ddot{y_L} = m_L \, g - \frac{y_L}{l} \, T, \\
    m_C \, \ddot{x_C} &= \frac{x_L-x_C}{l} \, T - u, \qquad
    (x_L - x_C)^2 + y_L^2 = l^2,
\end{aligned}
\label{eq:crane}
\end{equation}
where $T$ denotes rope tension and $x_L$, $y_L$, and $x_C$ are positions of the load and crane.

    \item \texttt{MultilinkControl\_DAE.fmu}: Multilink rear suspension corner model.
    See \autoref{fig:fmiopt_multilink_schematic} for a schematic and description.
    A corresponding ODE-FMU with the regular Model Exchange interface was also generated for comparison on the optimization task in \autoref{sec:fmiopt_experiment}.

\end{itemize}

Moreoever, a new FMU has been added to the list of Reference FMUs \footnote{
    GitHub repository: \href{https://github.com/modelica/Reference-FMUs}{https://github.com/modelica/Reference-FMUs}
} that conforms to the current version of \emph{fmi-ls-dae}. This is the \texttt{Roberts.fmu} which implements a problem from chemical kinetics, due to Robertson \cite{Robertson1966}. It implements the following equations in \eqref{eq:robertson-fmu}.
\begin{equation}
\label{eq:robertson-fmu}
\begin{aligned}
\frac{dy_1}{dt} &= -0.04\,y_1 + 10^4 y_2 y_3, \\
\frac{dy_2}{dt} &= 0.04\,y_1 - 10^4 y_2 y_3 - 3\times10^7 y_2^2, \\
0 &= y_1 + y_2 + y_3 - 1,
\end{aligned}
\end{equation}

with initial conditions
\begin{equation}
y_1(0) = 1, \qquad y_2(0) = y_3(0) = 0.
\end{equation}

\subsection{Simcenter Twin Activate - Simulation}
\label{sec:simcenter_experiment}
\begin{figure}[htb]
    \centering    \includegraphics[width=0.5\columnwidth]{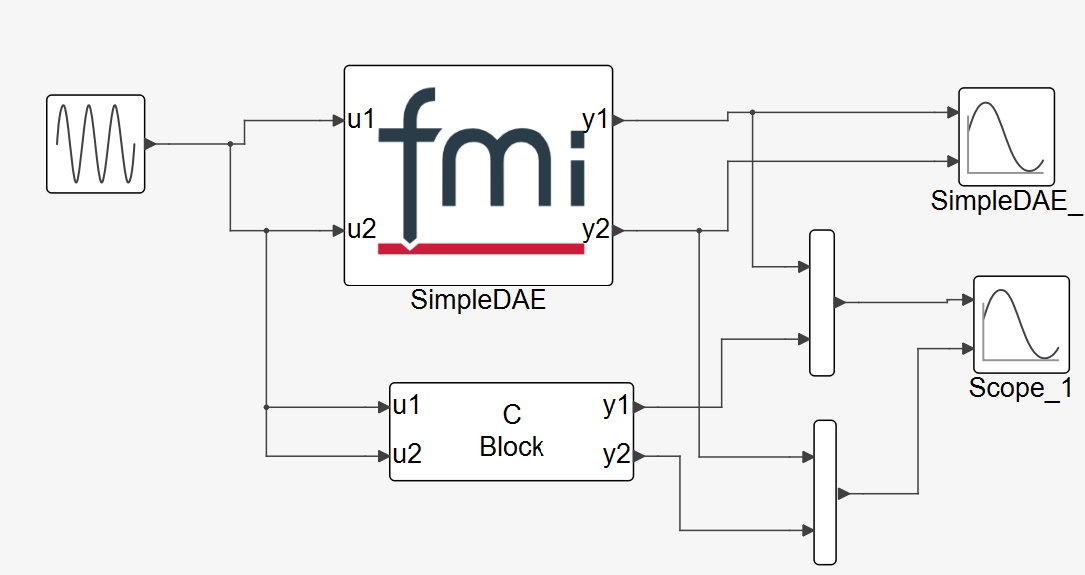}
    \caption{The simpleDAE FMU imported and a built-in block in Simcenter Twin Activate.}
    \label{fig:simpleDAE_ta_model}
\end{figure}

The two DAE-based FMUs \texttt{SimpleDAE.fmu} and \texttt{Fourbar1.fmu} were imported and tested.
As shown in \autoref{fig:simpleDAE_ta_model}, the internal DAE equations from the FMU were independently implemented in a built-in block to enable direct comparison and validation of the results.

\begin{figure}[htb]
    \centering
    \includegraphics[width=\columnwidth]{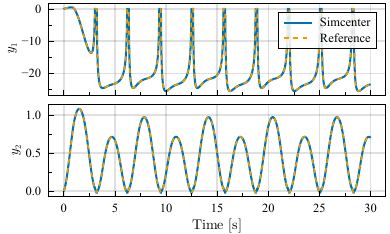}
    \caption{Simcenter Twin Activate simulation results for \texttt{SimpleDAE.fmu} with input $u_1(t) = u_2(t) = \sin(t)$ compared to reference solution performed by OpenModelica: outputs~$y_1$ (top) and~$y_2$ (bottom).}
    \label{fig:simpleDAE_ta_res}
\end{figure}

\noindent
\autoref{fig:simpleDAE_ta_res} presents a comparison of the simulation results from both the built-in block and the FMU, demonstrating a perfect match between the two implementations.

 \begin{figure}[htb]
    \centering    \includegraphics[width=0.5\columnwidth]{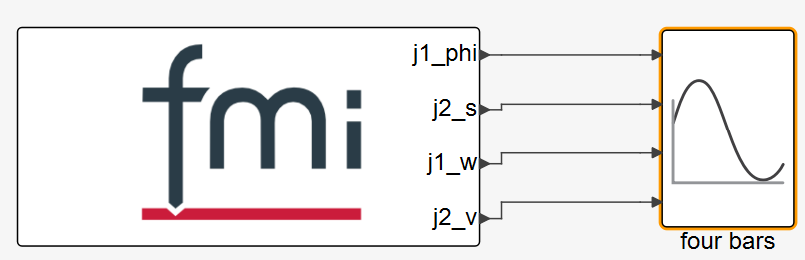}
    \caption{The fourbar1 FMU imported in Simcenter Twin Activate.}
    \label{fig:fourbar1_ta_model}
\end{figure}

\begin{figure}[htb]
    \centering
    \includegraphics[width=\columnwidth]{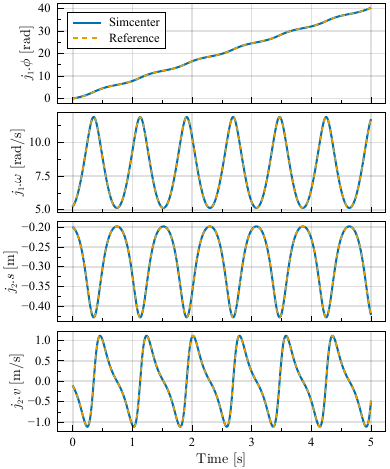}
    \caption{Simcenter Twin Activate simulation results for \texttt{Fourbar1.fmu} compared to the Modelica reference solution:
        revolute joint angle~$\varphi_1$, angular velocity~$\dot{\varphi}_1$, prismatic joint displacement~$s_2$, and prismatic joint velocity~$\dot{s}_2$.}
    \label{fig:fourbar1_ta_res}
\end{figure}

\noindent
\autoref{fig:fourbar1_ta_res} displays the simulation results of the FMU integrated into the model shown in \autoref{fig:fourbar1_ta_model}, which are identical to those produced by both Dymola and OpenModelica with the original Modelica model.

\subsection{FMIOPT - Dynamic Optimization}
\label{sec:fmiopt_experiment}

\begin{figure}[htb]
    \centering
    \includegraphics[width=0.9\columnwidth,clip,trim=100 100 100 100]{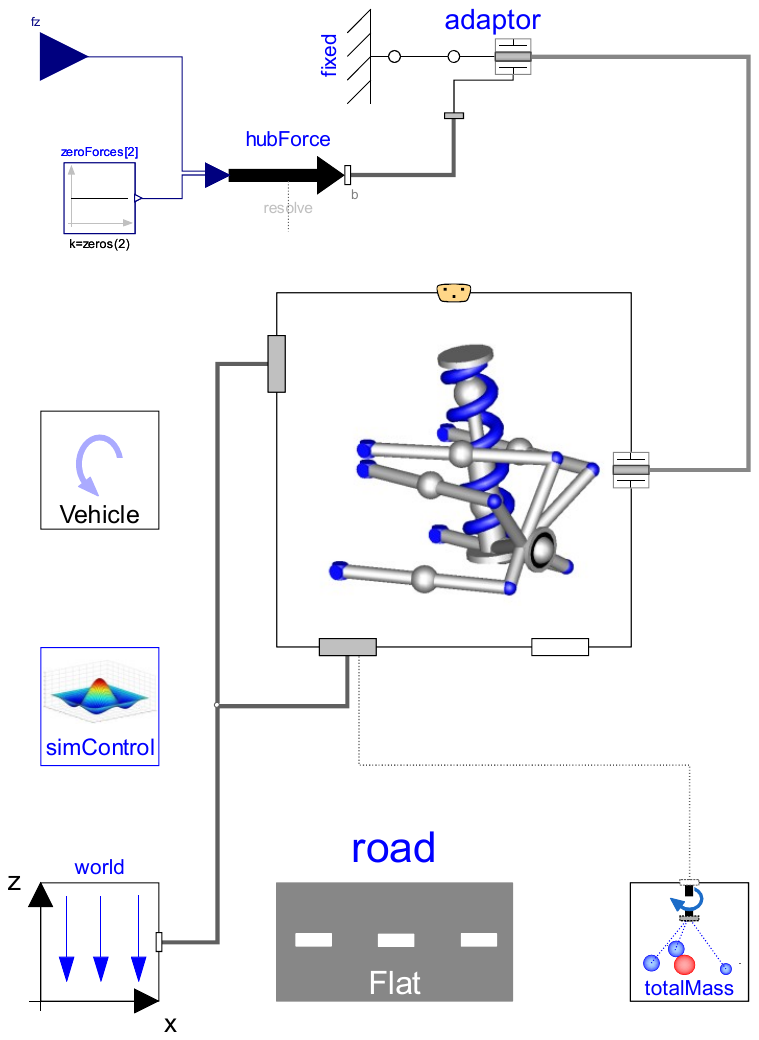}
    \caption{Multilink rear suspension corner model.
    The actuator force $f_z$ is the control input and the joint angle $\phi$ is the tracked output.}
    \label{fig:fmiopt_multilink_schematic}
\end{figure}

To exercise the layered standard on an industrially relevant model, we consider a multibody model of a rear multilink suspension corner.
This model was derived from the Claytex \texttt{Suspensions} library and exported as the FMU \texttt{MultilinkControl\_DAE.fmu} mentioned in the beginning of this section.
A schematic of the model is shown in \autoref{fig:fmiopt_multilink_schematic}.

After the kinematic constraints of the linkage are accounted for, the suspension corner has a single mechanical degree of freedom, yielding two differential states $\phi$ and $\omega = \dot{\phi}$.
A vertical force $f_z$ is applied at the wheel hub and the lower control arm joint angle $\phi$ is tracked against a sinusoidal reference.
This setup is similar to a vertical wheel bump experiment used to characterize suspension kinematics and the wheel rate.
However, rather than imposing the motion quasi-statically, the optimal control problem inverts the suspension dynamics to find the actuator force that produces a prescribed angular trajectory.

The model contains a kinematic loop that cannot be reduced to an ODE analytically, which results in a non-linear system of equations that must be solved at each time step during simulation.
This makes it an ideal candidate for testing the layered standard for DAEs.
Moreover, in SI units the model is poorly scaled, with forces at the wheel hub on the order of $10^3$--$10^4$~N and angles on the order of $10^{-1}$~rad, which can lead to numerical issues especially if the algebraics are not exposed to the optimizer, leading to a Newton-over-Newton solve that can produce noisy derivatives and failures in the inner Newton iteration.

After symbolic processing in Dymola, the model results in a semi-explicit index-1 DAE of the form \eqref{eq:ind1-dae} with two differential states ($\phi$ and $\omega$), a vector $\bm{z}$ of algebraic iteration variables originating from the kinematic loops of the suspension, and the scalar control input $f_z$.
The model is then exported as LS-DAE FMU using the \emph{fmi-ls-dae} export described above, and as a standard ODE FMU 3.0 for comparison.
The optimal control problem is set up as follows:
\begin{equation}
\begin{aligned}
    \min_{\bm{x}(\cdot),\,\bm{z}(\cdot),\,f_z(\cdot)} \quad
        & J = \int_0^T \!\!\left( w_\phi\,(\phi-\phi_{\mathrm{ref}})^2 + w_u\,f_z^2 \right) dt \\
    \text{s.t.}\quad
        & \dot{\bm{x}} = \bm{f}\!\big(\bm{x},\bm{z},f_z,t\big), \\
        & \bm{0} = \bm{g}\!\big(\bm{x},\bm{z},f_z,t\big), \\
        & \bm{x} = [\phi,\,\omega]^\top, \\
        & \phi(0) = \phi_0, \quad \omega(0) = \omega_0, \\
        & f_{z,\min} \le f_z(t) \le f_{z,\max},
\end{aligned}
\label{eq:fmiopt_ocp}
\end{equation}
on the fixed horizon $t \in [0,\,T]$ with $T = 2$~s.
The reference trajectory and consistent initial conditions are
\begin{equation}
\begin{aligned}
    \phi_{\mathrm{ref}}(t) &= \phi_0 + A_\phi\,\sin(2\pi f t), \\
    \omega_0 &= 2\pi f A_\phi,
\end{aligned}
\end{equation}
with amplitude $\phi_0=-0.28$~rad, $A_\phi = 0.13$~rad, and frequency $f = 1$~Hz.
The actuator bounds $f_{z,\min} = -f_{z,\max} = -2 \cdot 10^6$~N are intentionally loose and act as non-binding numerical safeguards rather than as physical limits.
The cost weights $w_\phi = 10^3$ and $w_u = 10^{-12}$ follow a Bryson-style normalization: with $\phi$ on the order of $10^{-1}$~rad and the optimal $f_z$ on the order of $10^3$--$10^4$~N, the two terms in the integrand are brought to comparable magnitude, with tracking dominating slightly so that $f_z$ is regularized only mildly.

The OCP is transcribed by direct collocation on $N = 150$ uniform intervals using a 3-point Radau IIA scheme.
On each collocation interval, the FMU is queried for $\bm{f}$, $\bm{g}$, and the corresponding first-order partial derivatives, and the algebraic variables $\bm{z}$ are introduced as additional decision variables.
The residual values $r = \bm{g}(\bm{x},\bm{z},f_z,t)$ exposed by the FMU through the layered standard are imposed as equality constraints $r = 0$ at every collocation node.
The resulting NLP is solved by IPOPT~\cite{Waechter2006} with the MA57 linear solver and a limited-memory BFGS Hessian approximation.

\begin{figure}[htb]
    \centering
    \includegraphics[width=\columnwidth]{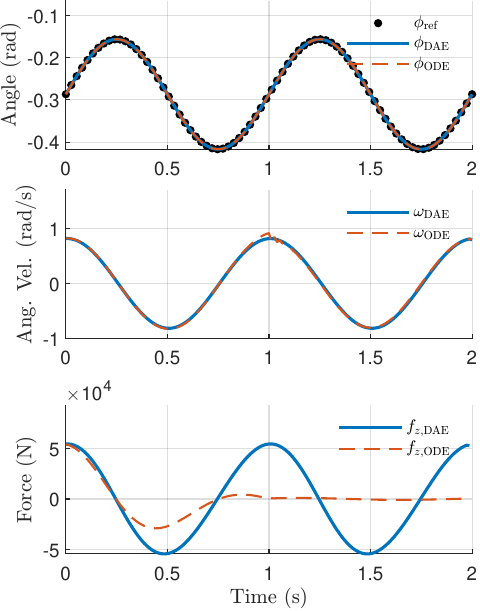}
    \caption{Optimal trajectories for the multilink suspension corner: tracked joint angle $\phi$ versus reference $\phi_{\mathrm{ref}}$, angular velocity $\omega$, and optimal actuator force $f_z$.}
    \label{fig:fmiopt_multilink_tracking}
\end{figure}

\autoref{fig:fmiopt_multilink_tracking} shows the optimal trajectories of the tracked angle, the corresponding angular velocity, and the optimal actuator force $f_z$ obtained from the DAE-FMU.
The reference is followed essentially exactly over the full horizon, while $f_z$ remains well within its prescribed bounds.

An attempt was also made to solve the same OCP using a regular (ODE) FMUs, where the kinematic loop is closed by the FMU-internal Newton iteration. Using the same transcription and solver settings, the ODE formulation run failed to converge, with IPOPT exiting with a \emph{Restoration Failed!} status after 369 iterations. The corresponding trajectories in \autoref{fig:fmiopt_multilink_tracking} are also clearly unphysical: the angular velocity $\omega$ exhibits spurious artifacts around $t = 1$~s and the optimal actuator force collapses to zero rather than tracking the reference. We caution against drawing broad conclusions about the failure of the ODE FMU formulation -- more investigation is needed to determine the root cause of the failure, which may be unrelated to the ODE vs. DAE formulation.

\subsection{MOO - Dynamic Optimization}
\begin{figure}[htb]
	\centering
	\includegraphics[width=\columnwidth]{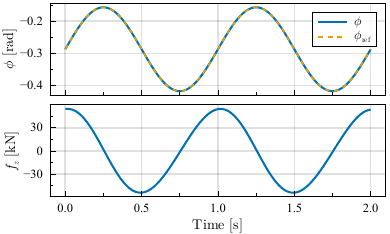}
	\caption{Solution of the multilink suspension corner problem solved using MOO.
    The joint angle $\phi$ closely follows the reference $\phi_\mathrm{ref}$ and the actuator force $f_z$ agrees with the DAE-FMU solution of FMIOPT.}
	\label{fig:moo_multilink_results}
\end{figure}

The DAE-FMU version of the multilink suspension corner OCP \eqref{eq:fmiopt_ocp} was additionally solved using MOO with the FMI interface described in \autoref{sec:moo_intro}.
The OCP is transcribed using $N = 400$ discretization points using direct collocation.
It is solved with IPOPT \cite{Waechter2006} and MUMPS \cite{Amestoy2001} chosen as linear solver, and converges to the optimum in 57 NLP iterations.
\autoref{fig:moo_multilink_results} shows the optimal joint angle $\phi$ against the reference $\phi_\mathrm{ref}$ as well as the actuator force $f_z$.

\subsection{CasADi - Optimal control, FMU export}
The example \href{https://github.com/casadi/casadi/blob/main/docs/examples/python/fmu_collocation.py}{\texttt{fmu\_collocation.py}} was added to CasADi's example collection, demonstrating how to find the optimal control that takes the load of the crane in \autoref{eq:crane} to a desired position while minimizing a quadratic cost.
Using the generated \texttt{Crane.fmu}, it implements a direct collocation method, solved with IPOPT \cite{Waechter2006} with first and second-order derivatives provided by CasADi.
The same example also demonstrates how an equivalent FMU can be exported from CasADi.

\subsection{fmusim - Simulating the Roberts FMU}
\begin{figure}[htb]
	\centering
	\includegraphics[width=\columnwidth]{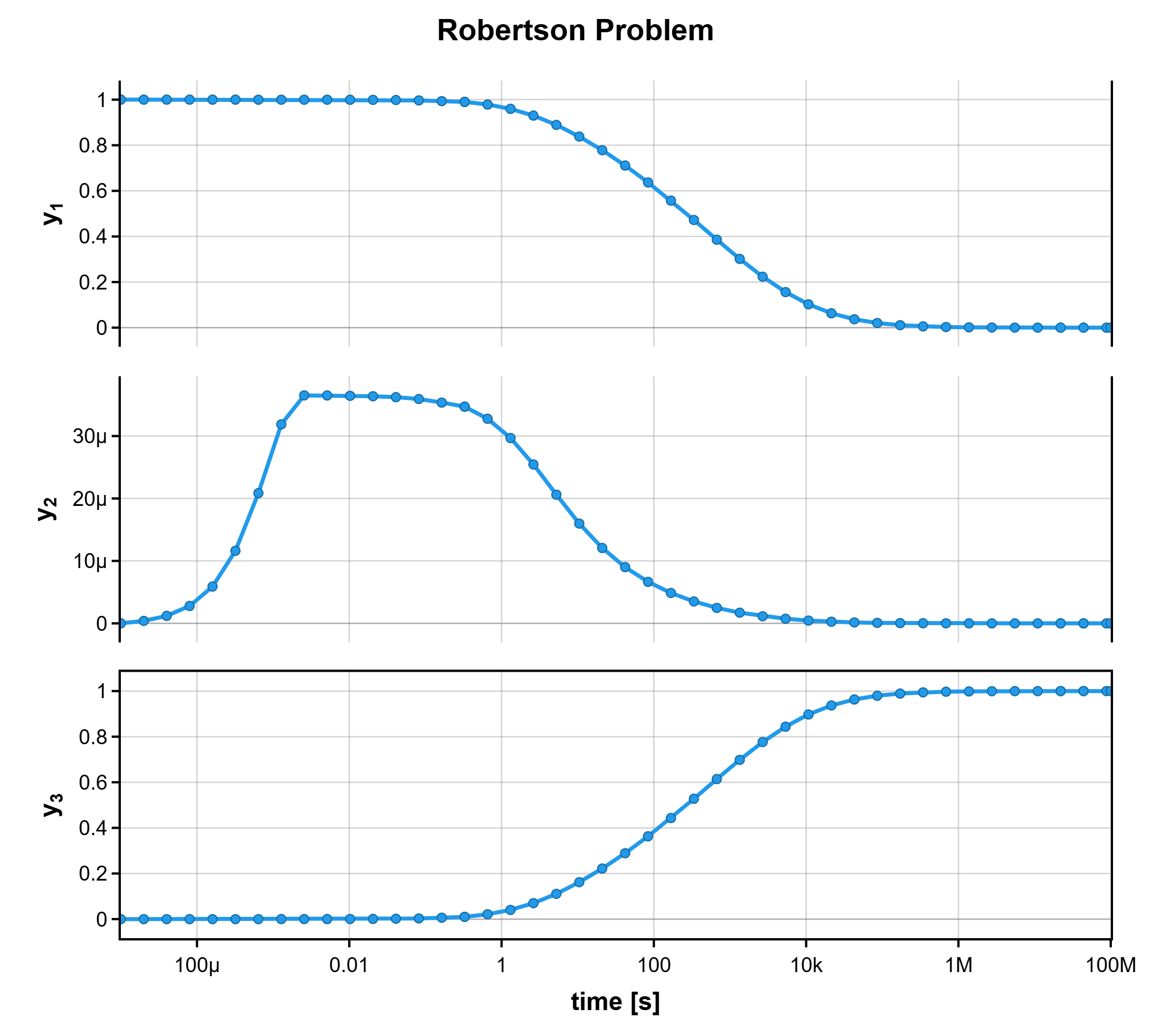}
	\caption{Solution of the \texttt{Roberts.fmu} simulated with fmusim, plotting the two states $y_1$ and $y_2$, and the algebraic variable $y_3$.}
	\label{fig:fmusim-roberts}
\end{figure}

The \verb|Roberts.fmu| can be imported and simulated as an \emph{fmi-ls-dae} FMU (using the default experiment in the FMU) in fmusim for example with the following command line arguments:

\begin{lstlisting}[
    breaklines=true,
    breakatwhitespace=true,
    language={}
]
fmusim simulate Roberts.fmu --enable-dae --solver ida --interface-type me --log-time-scale --show-plot --show-markers --output-interval 2
\end{lstlisting}

For a full list of possible options, one can run \verb|fmusim --help| and \verb|fmusim simulate --help|.
Running the above commands results in \autoref{fig:fmusim-roberts} which shows the two states and one algebraic variable over the time horizon defined by the default experiment in the FMU.

\section{Outlook} \label{sec:outlook}
This article focuses on exposing the DAE formulation for continuous time simulation, which will be included for the pre-release version \texttt{1.0.0-alpha.1}, planned for 2026.
The details of handling initialization and event iteration are under active development within the \emph{fmi-ls-dae} working group and are outlined in \autoref{sec:outlook-initialization} and \autoref{sec:outlook-events}.
Once these are finalized, the next milestone in development\footnote{
    GitHub: modelica/fmi-ls-dae milestone \href{https://github.com/modelica/fmi-ls-dae/milestone/1}{Alpha-2}
}
is the publication of pre-release version \texttt{1.0.0-alpha.2}, which is also planned for 2026.
\autoref{sec:outlook-higher-index} provides an outlook on potential support for higher-index DAE-FMUs in a later major release of the layered standard.

\subsection{Consistent Initial Condition}
\label{sec:outlook-initialization}
In addition to the discussed initialization methods from \autoref{sec:initialization} some simulation environments also allow other methods for index-1 DAEs, depending on the available information and the modeling intent.

Partially inconsistent initial conditions apply a correction or projection step to compute a
nearby consistent state before starting the integration.
In addition, a DAE may be initialized in steady state by solving the system under the assumption that all time derivatives of the differential states are zero,
yielding an equilibrium point that satisfies both the differential and algebraic equations.

Although the exact approach has not yet been finalized, the initialization type could be specified by selecting appropriate variables in the \verb|<InitialUnknown>| element.

\subsection{Event Handling}
\label{sec:outlook-events}

Another important aspect for the final release of \emph{fmi-ls-dae} is event handling and the associated reinitialization procedure.
Events may result from switching conditions, zero-crossing detections, synchronous events, or explicit state jumps.
Such events can invalidate the consistency conditions of the DAE and therefore require reinitialization before integration can continue.
After applying any prescribed jumps to the differential states, the algebraic variables must be recomputed such that the algebraic constraints are satisfied.
Depending on the model, this may require solving a non-linear system to establish a new consistent operating point.
Model invariants and conservation laws should also be enforced during this process.
Reinitialization takes place during event mode in cooperation with the importing simulation environment.

The current \emph{fmi-ls-dae} specification does not yet define a standardized mechanism for event-driven reinitialization, and
the procedure is therefore delegated to the importer and its underlying DAE solver using the information provided by the FMU.
Defining a portable and solver-independent reinitialization workflow remains an important topic for future standardization efforts.

Dymola's (non-FMI) DAE-Mode localizes the event time efficiently and accurately based on the ODE representation of the DAE \cite{Henningsson2019DAE}.
OpenModelica's (non-FMI) DAE-Mode requires the causalized ODE representation to handle synchronous events and discrete variables during event iteration \cite{BraunOpenModelicaDAE}.
Other tools might only use their (non-FMI) DAE-Mode for all of the event handling.
The authors believe that with the possibility to switch between DAE- and ODE-Mode during Reconfiguration Mode all of these cases can be incorporated by \emph{fmi-ls-dae}.
Development of reference DAE-FMUs showcasing event handling is planned for the first release of \emph{fmi-ls-dae}.

\subsection{Higher-Index Formulations}
\label{sec:outlook-higher-index}
This article focuses on the specific use case of exposing algebraic equations in an index-1 semi-explicit formulation, but the end goal of the layered standard is to support higher-index DAEs as well.
Part of the ongoing discussion at the time of writing is to include a standardized way to specify \emph{invariants} for fmi-ls-dae. Although they are not directly related to higher-index problems, one idea is to use the proposed \verb|<Invariant>| element, such that by letting the element contain a \verb|derivative| attribute pointing to a \verb|<Residual>| element, it could define the higher-index residuals.
An importing tool may use this structural information to apply, for example, index-reduction techniques or drift-correction methods.

\section{Conclusion}
We proposed \emph{fmi-ls-dae}, a layered standard on top of FMI 3.0 Model Exchange that exposes algebraic equations of a model and their associated algebraic variables as part of a semi-explicit index-1 DAE, together with the extensions to the FMI XML schema required to describe the residuals and algebraic variables in a backwards-compatible way.
With this extension we demonstrated the standard end-to-end through prototype implementations in five different tools: Dymola and CasADi were extended to export DAE-FMUs conforming to the layered standard, and CasADi, FMIOPT, Simcenter Twin Activate, fmusim, and MOO were extended to import them for simulation and dynamic optimization.
For one industrially relevant multilink rear suspension corner model, the DAE-FMU formulation was able to converge to the optimal solution, while the equivalent ODE-FMU formulation failed to converge.
\section*{Author Contributions}

All authors contributed to the specification through collaborative meetings and discussions, and all authors have contributed to the writing of the paper.

\begin{itemize}
    \item The paper was coordinated by Elmir Nahodovic, Andreas Heuermann, and Joel A.\ E.\ Andersson.
    \item Andreas Heuermann and Joel A.\ E.\ Andersson served as working group leaders.
    \item Elmir Nahodovic led the index-1 semi-explicit DAE specification and developed the Dymola prototype for this proposal, with technical support from Erik Henningsson and Hans Olsson.
    \item Joel A.\ E.\ Andersson designed and implemented the CasADi and FMIOPT prototypes.
    \item Adwait Verulkar and Srikanth Sivaramakrishnan re-initiated the \emph{fmi-ls-dae} project by proposing it in the FMI Design Meetings and provided the industrial Multilink study with FMIOPT.
    \item Masoud Najafi extended Simcenter Twin Activate to support the import of such FMUs.
    \item Linus Langenkamp developed the prototype with MOO, with support from his supervisor Bernhard Bachmann. Andreas Heuermann supported in the protoype for MOO.
    \item Christian Bertsch, as the FMI project leader, contributed with his knowledge of layered standards for FMI.
    \item Torsten Sommer contributed by extending his command line tool fmusim to import \emph{fmi-ls-dae} FMUs, and with the new \verb|Roberts.fmu| in the Reference FMUs.
\end{itemize}

\section*{Acknowledgements}
Part of this work was conducted within the EUREKA ITEA4 OpenSCALING project (\href{www.openscaling.org}{openscaling.org}) by the German Federal Ministry of Education and Research (BMBF) under grant number 16IS23062E and the Swedish Innovation Agency (grant number 2023-00969).
The authors would like to express their sincere appreciation to the OpenSCALING project for their support, collaboration, and shared commitment in advancing open standards for modeling and simulation technologies.

We want to thank the FMI Project members, especially also the current and former members of the DAE working group!

\printbibliography

\end{document}